# A Comparative Study of Statistical Learning and Adaptive Learning

Ayan Roy[1], Kaustuvi Basu[2]

[1]Post Graduate Student, St. Xavier's College, Kolkata, ayan.199316@gmail.com
[2]Post Graduate Student, St. Xavier's College, Kolkata, basu.kaustuvi@gmail.com

## Abstract

**Numerous strategies have been adopted in order to make the process of learning simple, efficient and within less amount of time.. Classroom learning is slowly replaced by E-learning and M- learning. These techniques involve the usage of computers, smart phones and tablets for the process of learning. Learning from the internet has become popular among the e-learners where learner tends to rely greatly upon information provided by the World Wide Web. However, the e-learners have to go through a huge volume of data produced by the first tier search engine, some of which are not suited to the interest of the user. Various strategies, namely Statistical Learning and Adaptive Learning, have been adopted to cater to the need of the user and produce data best suited to the interest of the user. The authors have tried to present a comparative study of Statistical Learning and Adaptive Learning based on certain parameters, which arise from the characteristics of the learning process. As a consequence of the comparative study, it has been concluded that Adaptive learning is more efficient than Statistical learning.**

## 1. Introduction

Intelligent e-Learning Environments (ILE) enhance the concept of e-learning systems. It improves the teaching efficiency, by adapting to each Learner Profile (LP), and by providing multiple supports to the tutor. These components guide the trainee through the learning process, offer a platform for co-operative learning and knowledge discovery, and customize the presentation to learner's preferences, interests and needs. Artificial intelligence and evolutionary tools are used to this end.

Data available in the internet is quite enormous. Some of the data available can be put into use by the user while some are rendered inappropriate to their requirements. While a user wants to find some data using the World Wide Web, the data presented by the search engine may be not in accordance to the interest of the user. However, it becomes difficult for the user to choose the information appropriate to their interest as the data presented by the search engine is huge and finding the relevant information is time consuming.

Below is just a diagrammatic representation of the concept involved in the e-learning system.

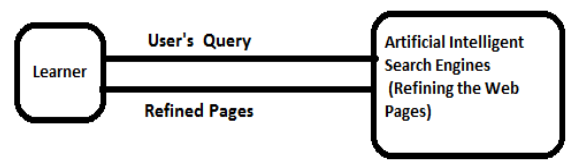

**Fig: 1: Block Diagrammatic representation of e-learning**

According to the model the query provided by the user is passed on to several agents present within the system. The agent considers the query provided by the user to have knowledge about the actual requirement of the user. Base on this fact, the results are further filtered and provided to the user.

There are various methods of learning. Out of them two most important methods are

- **E-Learning** – It is a strategy which involves learning our courses in our personal computers or laptops. It is not much portable and it limits the users choices of learning in at any place. It involves learning from any device dependent upon the actions of electronics, such as television, computers, microcomputers, videodiscs, video games, cable, radio interactive cable, videotexts, tele text, and all the other devices in the process of being invented that are electronic in nature [13].

It is formal and structured. It is way more comprehensive. Here the lesson is not broken up into micro lessons but are broken up in the form of chapters. The chapters by itself are comprehensive. It can be accessed in laptops and computers, therefore making it mandatory to have a static environment. The amount of time taken to cover a particular topic may range from 20-45 minutes.

➢ **Mobile Learning**– It is a refined version of E-Learning. As mentioned E-Learning curbs the idea of portability. However this idea has been taken into consideration and m-learning has paved its way to simplify the process of e-learning. This can be used to anywhere to continue the learning process and finds its usefulness mostly among the students. The intersection of mobile computing (the application of small, portable, and wireless computing and communication devices) and e-learning (learning facilitated and supported through the use of information and communications technology) [14].

It is much more informal than e-learning strategy and can be structured as well as unstructured. Here the entire lesson is broken up into micro-lessons, thus enabling us to access it whenever required. It requires the usage of cell phones or any relevant mobile devices and hence can be accessed anywhere and everywhere. The amount of time required to cover a particular topic is approximately 4-5 minutes.

Later in this paper we have discussed the various approaches that has been adopted which helps to provide the desired results to the learners based on their criteria and a comparative study of the two approaches with reference to certain defined parameter. This paper mainly highlights the difference between the two learning strategies and the advantage that each strategy holds over the other. We have also highlighted the necessities of the defined parameters and its effect upon the learning experience of the learner. We have also mentioned the various ways the strategies works with proper block diagrams and suitable expressions.

## 2. Related Research

Prominent Researchers have evolved in the field of E-Learning to broadcast the various techniques adopted by the various researchers to produce the personalized pages displayed by the search engines. These techniques are adopted in order to improve the search quality and focus the search based upon the interest of the e-learner. These may vary from person and person and it is upon the model to adapt itself to cater the needs of the e-learner.

I.Seher has proposed a model which would extract the words in the neighbourhood of a target word and connecting it with the surroundings of other occurrences of the same words in the text. [1]To get the result of expanded Query there are plenty of search engines are available but first tier search engines like Google, Yahoo and Bing are providing full proof result compared to others.[16]

Mirjam Kock of Johanes Kepler University worked in the field of adaptive learning to broadcast the vital differences between the statistical learning method and intelligent learning technique. Furthermore, he has compared the behaviour and performance of

different classification algorithms based on real data usage. [2]

Axita Shah and Sonal Jain, in their model, passes the results of the first tier search engine through a set of agents (Query Expansion Agent, Middleware Filtering Agent etc) to produce the personalised Search Engine pages best suited to the interest of the e-learner. They had therefore refined the results of the first tier search engines to produce an output desirable to the user. [3]

Barbara Oakley, Rebecca Brent, Richard M. Felder, and ImadElhajj in their model focus on data driven modelling of student's interaction, prediction of a student to answer correctly and whether it is beneficial to students in terms of learning. [4]

Web Recommender system has been proposed which uses advanced feedback mechanism to extract the information best suited to the interest of the user and the learning behaviour of user [5], learning objects recommended for the users are obtained using suitable strategies that is solely based upon content based filtering and collaborative filtering approaches. [6]

Due to the large amount of irrelevant information presented by the first tier system, the original web cannot be directly used in web mining procedure. To achieve suitable data mining the design of data e-learning web house has been proposed. [7]Techniques have also been provided by researchers in which the interest of the user is automatically learned from the web usage data using the web mining strategy. [8]

Hafidh Ba Omar through his model improved personalized learning using learning style. He had used two techniques for identifying learning patterns of
Learners and sequence of choosing learning resources: [9]

Research work has laid a foundation for mining, tracking, and validating evolving multifaceted user profiles on Web sites that have all the challenging aspects of
real-life Web usage mining, including evolving user profiles and access patterns, dynamic Web pages, and external data describing the Web content [10],[11].

Edward J.Cherian and Paul Williams highlighted the advantages that the mobile learning has over e-learning and the various ways in which the disadvantages of e-learning are overcome by mobile learning system. [15]. He focussed on the relationship between mobile learning, e-learning and the digital learning strategies and defined various ways in which mobile learning proves to be dominant over the other strategies.

## 3. Intelligent vs. Statistical Approach

The mode of learning should be appropriate enough so that it becomes more users friendly. It should ease the efforts of the e-learners and make it a pleasurable one. In this section we have solely emphasized on two important learning techniques, Statistical Method and Adaptive (Intelligent) method. These two techniques have played a vital role over the years in order to develop the concept of e-learning and m-learning. The two methods are different from each other, former based on pure statistics while the later based on present situation of the learner. The Statistical method was the first important step taken towards this field. However, it had a few shortcomings because of which Adaptive learning strategy was established. It overcomes the disadvantages of the Statistical approach

### 3.1 Adaptive Learning (Intelligent Method)

An important application of the intelligent E-Learning System is the Adaptive E-Learning. This is a system which is mostly used among students in order to determine the student's level of knowledge. It is enriched with educational content and provides unique feedback to the students based on their knowledge

It is a system which helps to educate students by incorporating the various talents of the educators. It helps to serve the students according their varying level of knowledge and individual needs other student. As different students have different ways of learning, it adapts itself according to the needs of the student and provides feedback based on it. Thus, it is a personalised educational tool for any student which helps to improve the performance of the student and aims to curb out any possible failures and pointing out the weakness of the student through repeated experimentation and analysis. This also involves pointing out the flaws of a certain student through repeated examinations.

Fig: 2 typically illustrates one such example where Adaptive E-Learning plays a vital role in the performance of student.

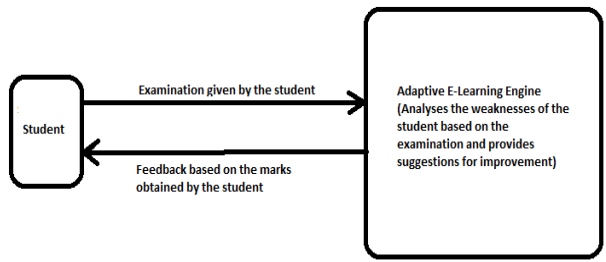

**Fig: 2: Block Diagrammatic representation of Adaptive Learning**

According to this model, the weakness of the student is revealed by a series of examinations conducted by the system. The student appears for the examination and writes the examination according to his knowledge. Based upon the answers, the system evaluates the particular topic that needs to be improved and provides it as a feedback to the student. It also describes the various ways that needs to be adapted in order to improve on those particular areas. By this technique the student gets to analyze his weaknesses using a tutor like experience digitally

### 3.2 Statistical Method

This is a method which is based on pure statistics that is the previous record of the user. It uses statistical formula to determine the interest level of the user. It considers the various aspects of the learner before laying it as a basis for future prediction. For example it considers the read activities, to determine the various web pages that are visited by the user and based on it the statistical metrics are used to determine the interest level of the learner. The various statistical metrics used are standard deviation, mean, variance etc.

A statistical variable is inflexible even if it contains variable elements. It means that the core does not change for different scenarios. This (simplified) approach can be improved, e.g. by adding weights, and in theory this improved version might be sufficient, but it still carries some non-obvious risks.[12]   A statistical formula, even if it contains variable elements, is inflexible, meaning that the core does not change for different scenarios.
The following equations represent the various statistical formulas which are used:-

$$X' \text{ (mean of n numbers)} = \frac{x1+x2+x3...+xn}{n}$$

Where n=total number of elements, x1 to xn = constants

$$\delta^2 = \sum_{i=1}^{n}(xi - x')^2$$

Where $\delta^2$ = Standard Deviation of n elements
x'= mean of n elements

This process is way more rigid as it does not have the capability to adjust itself according the user's behaviour. There can be courses where communication plays a more important role than educational content, and users might differ in their communication and learning behaviour in several ways. This process is based upon a set of pre defined process and procedures. Therefore, it adds to inflexibility to the entire process

## 4. Comparative Study Between Statistical and Adaptive Learning

In this section we have included a comparative study of the above discussed learning techniques based upon the following set of parameters:-

- **Accuracy** – The model should be accurate enough so that it clearly highlights the drawbacks of the users and the areas which demands improvements without ambiguity.
- **Flexibility** – The model should be flexible enough. It should not be built based upon a set of rules and procedures and any changes in the model would require no additional effort or change of the model.
- **Adaptability** – The model should be adaptable. This means that the system or rather the model should be able to adapt according to the needs of the user. This is needed particularly to highlight the individual needs of the e-learners as it may differ from person to person.
- **Rigidity** – This is needed in order to ensure adaptability of the system. The model should not be built upon a set of predefined rules and procedures and should be able to adjust accordingly in order to cater the needs of the e-learners.

| Parameters | Statistical Learning | Adaptive Learning |
|---|---|---|
| Approach | It is formal and structured | It is way more informal than e-learning. It can be structured as well as unstructured |
| Methodology | It is way more comprehensive. It does not involve breaking up into micro lessons. It does covers the topic in the form of modules and chapters which are no less comprehensive by itself | Here the entire contents are broken down into micro lessons, enabling the learner to access it whenever required. |
| Accuracy | This process is based on user's previous record and statistic, hence often rendered inaccurate | It is based on the recent knowledge of the user, hence making it a reliable one |
| Flexibility | It works for certain scenarios but it turns out to be inflexible towards others. | It is much more flexible |
| Adaptability | It does not have the capability to adapt according to the requirement of the user. | It adapts itself according to the unique requirement of the user. |
| Rigidity | It is rigid as it based upon a core concept based on which the process works. | It is not rigid as it adapts according to the behaviour of the user and there are no pre defined concept for the process |
| Devices | Used in laptops or personal computers | Used in mobile devices (cell phones, tablets) |
| Access | Can be accessed only at the computer desks or at a static environment. | Can be accessed relatively anywhere and everywhere as they are portable. |
| Time spent | It may involve 20-45 minutes to cover a particular topic | The amount of time spent to cover a topic is approximately 4-5 minutes |

**Table 1: Brief Comparison between Statistical Learning and Adaptive Learning**

The specified parameter have been chosen because these parameters play a vital role in making the entire concept user friendly and makes the learning process appear effortless to the user. User tends to look for a learning process which is hassle free, accessible from anywhere and which caters to their interests.

A detailed analysis of the comparison suggests that Statistical Learning is way more comprehensive and structural than Adaptive Learning is a bit low compared to Adaptive Learning strategy. On the other hand, Adaptive learning strategy is much more user friendly as it adapts itself to the needs of the user. Moreover it is much more portable, less rigid and much more flexible than the Statistical Learning strategy. The efficiency of Adaptive Learning is also better compared to that of the Statistical Learning Strategy.

## 5. Conclusion and Future Scope

The main purpose of this paper is to illustrate the various learning strategies adopted to suit the interest of the e-learners and the difference between Adaptive Learning and Statistical Learning. However the paper further aims to highlight the advantages and the disadvantages that each of the model holds over the other. It is a well proven fact that the Adaptive learning approach has responded quite well compared to that of the Statistical approach. Many researches has been held and are still conducted in the field of adaptive learning to make the entire process more user friendly and accurate. The Adaptive Learning strategy responds well when learning about the human behaviour is concerned and has so far yielded much efficient result compared to that of the old traditional Statistical method. Many models have been designed so far which effectively adapted itself according to the specific requirements of the user. The intelligent method is extensible in many ways. It can be used for building models pertaining to individual users, groups or clusters. These models however greatly contribute to learning through individual experience and personal explorations.

**Authors' Profile**

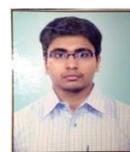

**Ayan Roy** is a post graduate student in Computer Science at St. Xavier's College, Kolkata. His current research includes works in the fields of Mobile Learning, Data Mining and Cryptography

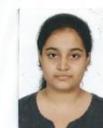

**Kaustuvi Basu** is a post graduate student in Computer Science at St. Xavier's College, Kolkata. She is interested in research work in the fields of Green Computing in Higher Education, Mobile Learning and Data Mining.